\documentstyle[aps,eqsecnum,epsf]{revtex}
\begin{document}
\draft

%%%%%%%%%%%%%%%%%%%%%%%%%%%%%%%%%%%%%%%%%%%%%%%%%%%%%%%%
\title{Parametrization of the octupole degrees of freedom}

\author{C. Wexler}

\address{Department of Physics, University of Florida, 
        Gainesville, Florida 32611\footnote{Current address.} \\ 
        and \\ 
        Departamento de F\'{\i}sica ``Juan Jos\'e Giambiagi,'' 
        Facultad de Ciencias Exactas y Naturales, 
        Universidad de Buenos Aires.\\ 
        Pabell\'on 1, Ciudad Universitaria, 
        (1428) Buenos Aires, Argentina} 

\author{G. G. Dussel\footnote{On leave of absence from the  
Comisi\'on Nacional de Energ\'{\i}a At\'omica.}\footnote{Member of the
Consejo Nacional de Investigaciones Cient\'{\i}ficas y T\'ecnicas.} }  

\address{Departamento de F\'{\i}sica ``Juan Jos\'e Giambiagi,'' 
        Facultad de Ciencias Exactas y Naturales, 
        Universidad de Buenos Aires.\\ 
        Pabell\'on 1, Ciudad Universitaria, 
        (1428) Buenos Aires, Argentina}

\date{April 1999}
\maketitle

%%%%%%%%%%%%%%%%%%%%%%%%%%%%%%%%%%%%%%%%%%%%%%%%%%%%%%%%
\begin{abstract}
A simple parametrization for the octupole collective variables is proposed
and the symmetries of the wave functions are discussed in terms of the
solutions corresponding to the vibrational limit.
\end{abstract}
%%%%%%%%%%%%%%%%%%%%%%%%%%%%%%%%%%%%%%%%%%%%%%%%%%%%%%%%

\thispagestyle{empty}

% -----------------------------------------------------------------
\section{Introduction}

\smallskip

The collective description of the octupole degrees of freedom has been a
long standing problem in nuclear physics\cite{Bu}. The phenomenological
description of low-energy vibrational states \cite{Bo52,Bo53} discussed
them in terms of simple ``surface modes'' thus allowing for the description
of the octupole degrees of freedom in terms of seven collective variables
$\alpha _{3\mu }$\cite{Bo52,da}. In the eighties, theoretical
calculations\cite{calcul} predicted the existence of octupole stable
deformations and this problem aroused considerable interest, especially in
the Ce-Ba and the Rn-Th regions. The level scheme of a few moderately- or
weakly- deformed nuclei, such as $^{64}Ge$\cite{Ennis},
$^{148}Sm$\cite{Urban},$^{218}Ra$\cite{Vanin} or $^{233,225}Ra$\cite{Chas}
presents features that may be related to octupole instabilities and
softness of the nucleus with respect to possible exotic octupole
deformations. Lately\cite{Cock} there has been evidence for the
existence of stable octupole deformations in the Rn-Th region making the
problem of the collective description of this degree of freedom more
actual. It has been shown that this type of collective excitations can
significantly change the fusion cross section for heavy ions \cite{Ca}.
Initially, only octupole deformations of the $Y_{30}$ type were considered,
but the possibility of other type of deformations (such as
$Y_{32}$\cite{Li}or $Y_{31}$\cite{Pie}) has also been discussed. The study of
more general structures was intended without much success \cite{Lipas68}. There
were also several proposals related to the appropriate parametrization of
the collective variables. Rohozinski\cite{Ro} proposed a parametrization
based on the symmetries coming from the $O_h$ group of transformations of
the frame of reference. Hamamoto\cite{ha} wrote octupole shapes in terms
of the irreducible representations of the octahedron group $A_1,F_1(k)$
and $F_2(k)$ and suggested that a parametrization that covers all possible
octupole deformed shapes (without double counting the same shape) is
provided by removing $F_2(k)$ (eliminating in this way three parameters). 

In the present paper we return to the idea used for the definition of the
intrinsic system in Bohr's paper\cite{Bo52} (also used by other authors
for the collective description of the pairing degrees of freedom): in the
intrinsic system the inertia tensor (related to the rotational part of the
kinetic energy) must be diagonal. 

In Sect.\ 2 we review some features of the collective excitations, the
Hamiltonian, we discuss how to separate the rotational and intrinsic
degrees of freedom and we propose a simple parametrization. We use the
Pauli prescription to quantize the quadratic Hamiltonian. In Sect. 3 we
discuss the structure of the wave functions and of the elementary scalars
or invariants (as defined in Ref.\ \cite{KB}), as well as the symmetries of
the wave function. In Sect.\ 4 we discuss the rigid rotor that illustrates
the differences between the rotational bands related with octupole
deformations as compared with the quadrupole ones\cite{DF}. Section 5 
summarizes the results obtained in the present paper. 

\vfill
\eject

\noindent

\section{Formulation of the problem}

\subsection{The collective variables}

The theory of collective oscillations has been developed long time ago by
several authors\cite{Bo52,Bo53,BM2}. The classical case corresponds to
quadrupole oscillations that were described in detail by A. Bohr in 1952.
A very clear and pedagogical discussion concerning the symmetries and wave
functions for this problem can be found in a paper\cite{KB} by K.  Kumar
and M. Baranger, where a numerical solution for this problem was first
proposed. A collective treatment was also performed for the pairing
degrees of freedom: the pairing acting only on one type of particles
yields a two dimensional collective description\cite{pair1}, while the
corresponding to the T = 1 case yields a 6 dimensional one\cite{pair2}. In
the three cases described, the recipe used to found the collective
description has the same basic ingredients: 

a) The kinetic energy was written explicitly in terms of the time
derivatives of the collective variables.

b) The collective variables (of dimension five, two or six, depending on
the case) were written in the intrinsic system, isolating the intrinsic
variables from those that describe the motion of the intrinsic system (the
three Euler angles in the five and six dimensional cases, one angle in the
two dimensional one). In all cases the intrinsic system was defined as the
system where the inertia tensor (the part related with collective
rotations) were diagonal. 

c) The kinetic energy was written in terms of the Euler angles and the
intrinsic variables, and the coupling between both types of degrees of
freedom was studied. Regarding the potential energy it has parts that
depend only on the collective variables (and not on their time
derivatives).  This part can be written in terms of elementary
scalars\cite{KB} that are constructed using the collective variables. For
example, in the quadrupole case there are only two of such scalars
($\beta ^2$ and $\beta ^3\cos (3\gamma )$, using Ref.\ \cite{Bo52}
notation), 
in the pairing case (between particles of one type) there is only one of
such scalars ($\Delta ^2$ using Ref.\cite {pair1} notation) and in the
T = 1 pairing case there are three of such scalars ($\Delta
^2,\,\Delta ^2e^{\pm 4i\phi }\cos (2\Gamma )$, using Ref.\cite {pair2}
notation). 

d) The general symmetry properties of the wave functions that are related
to different choices of collective (Euler) angles and intrinsic
variables corresponding to the same collective variables in the Lab.
system are used to define the range of the different intrinsic variables.
In some simple cases it was also possible to write down explicitly some
elementary tensors \cite{KB,gam,pair3} in terms of the collective
variables. 

In the present paper we will follow this program for the octupole surface
oscillations, relating them to the shape of the nucleus. Many times the
surface of the nucleus, in polar coordinates, has been expressed in terms
of the nuclear radius as

\begin{equation}
R(\theta ,\phi )=R_0(1+\sum_{\lambda \mu } \alpha_{\lambda ,\mu}
        Y_{\lambda ,\mu }(\theta ,\phi )) \,,
\end{equation}

\noindent
where $R_0$ is the radius of the nucleus in its spherical equilibrium
shape and $\alpha _{\lambda ,\mu }$ are the collective coordinates that
describe the deformation of the nuclear surface. We use for the
$Y_{\lambda ,\mu }$ the spherical harmonics satisfying the Condon and
Shortley convention, i.e. $ Y_{\lambda ,-\mu }=\left( -1\right) ^\mu
(Y_{\lambda ,\mu })^{*}$. As the radius has to be real, it follows that
$\alpha_{\lambda ,-\mu } = \left( -1\right)^\mu 
(\alpha _{\lambda ,\mu})^{*}.$

It is assumed that these variables change slowly with time, and therefore
it is usual to express their kinetic energy as a quadratic function of the
velocities as

\begin{equation}
{\cal T}=\displaystyle{\frac 1 2}
        \sum_{\lambda \mu}
        B_\lambda |\dot \alpha _{\lambda,\mu}|^2 \,,
\end{equation}

\noindent
where $\dot \alpha _\mu $ is the time derivative of $\alpha _\mu$. It is
well known that the coefficients $\alpha _{\lambda ,\mu }$ written in Lab.
system and the $\alpha _{\lambda ,\mu }$ written in the intrinsic system
(that we will denote by $a_{\lambda ,\mu }$ to avoid confusions) are
related by (in what follows we use the notation of Ref.\cite{BM2}) 

\begin{equation}
a_{\lambda ,\mu }=\sum_\nu {\cal D}_{\mu ,\nu }^\lambda (\omega _i)\alpha
_{\lambda ,\mu }  \label{alam} \,,
\end{equation}

\noindent
where $\omega _i$ are the three Euler angles. As usual the ${\cal D}_{\mu
,\nu }^\lambda (\omega _i)$ functions, which are related to the matrix
element of the rotation operation, are defined by

\begin{equation}
{\cal D}_{\mu ,\nu }^\lambda (\omega _i)=<\lambda ,\mu |\exp (i\psi I_z)\exp
(i\theta I_y)\exp (i\phi I_z)|\lambda ,\nu > \,,
\end{equation}

\noindent
where $\omega_i \equiv (\psi ,\theta ,\phi )$ are the Euler angles.

From Eq.(\ref{alam}) it follows that

\begin{equation}
\alpha _{\lambda ,\mu }=\sum_\nu [{\cal D}_{\mu ,\nu }^\lambda (\omega
_i)]^{*}a_{\lambda ,\nu } \,,
\end{equation}

\noindent
with $[{\cal D}_{\mu ,\nu }^\lambda (\omega _i)]^{*}=(-1)^{\nu -\mu
}{\cal D }_{-\mu ,-\nu }^\lambda (\omega _i)$. 

In order to evaluate the kinetic energy we must write down explicitly the
time derivative of $\alpha _{\lambda ,\mu }$:

\begin{equation}
\dot \alpha _{\lambda ,\mu }=\sum_\nu [{\cal D}_{\mu ,\nu }^\lambda (\omega
_i)]^{*}\dot a_{\lambda ,\nu }+\sum_\nu [{\dot {{\cal D}}}_{\mu ,\nu
}^\lambda (\omega _i)]^{*}a_{\lambda ,\nu }\,.
\end{equation}

\noindent
The time derivative of the ${\cal D}_{\mu ,\nu }^\lambda (\omega _i)$
functions can be expressed as

\begin{equation}
{\dot {{\cal D}}}_{\mu ,\nu }^\lambda (\omega _i)=i\sum_{m,\kappa }q_\kappa 
{\cal D}_{\mu ,m}^\lambda (\omega _i)(M_\kappa )_{\nu ,m}\,,
\end{equation}

\noindent
where $M_\kappa $ are the $(2\lambda +1)$ dimensional representation of
the angular momentum operator and satisfy the commutation relations

\begin{equation}
[M_1,M_2]=-iM_3 \,,
\label{conmu}
\end{equation}

\noindent
and cyclic permutations (note the signs). It is customary to use a
representation where $(M_3)_{m,n}=m\delta _{m,n}.$ The magnitudes $
q_\kappa $ are the angular velocity components in the $\kappa $ direction.
We can then express the kinetic energy as 

\begin{equation}
{\cal T} = \displaystyle{\frac B2} \sum_\mu (\dot 
        \alpha _{\lambda ,\mu })^{*} \dot \alpha_{\lambda ,\mu }
        = {\cal T}_{vibration}+{\cal T}_{rotation}+{\cal T}_{coupling} \,,
\end{equation}

\noindent
where for simplicity a single $\lambda$ is considered, and

\begin{eqnarray}
{\cal T}_{vibration}&=&\displaystyle{\frac B2}
        \sum_{\mu ,\nu ,\nu ^{^{\prime }}}{\cal D}_{\mu ,\nu}^{\lambda *}
        {\cal D}_{\mu ,\nu ^{^{\prime }}}^\lambda \dot 
        a_{\lambda,\nu ^{^{\prime }}}^{*}
        \dot a_{\lambda ,\nu }  \nonumber \\
{\cal T}_{rotation} &=& \displaystyle{\frac B2}
        \sum_{\mu ,\nu ,\nu ^{^{\prime }},m,m^{^{\prime}},
        \kappa ,\kappa ^{^{\prime }}}{\cal D}_{\mu ,m}^{\lambda *}
        {}^{*}{\cal D}_{\mu ,m^{^{\prime }}}^\lambda 
        a_{\lambda ,\nu }a_{\lambda ,\nu ^{^{\prime}}}^{*}
        q_\kappa q_{\kappa ^{^{\prime }}}(M_\kappa )_{\nu ,m}
        (M_{\kappa^{^{\prime }}})_{m^{^{\prime }},\nu ^{^{\prime }}}
        \\
{\cal T}_{coupling} &=& i\displaystyle{\frac B2}
        \sum_{\mu ,\nu ,\nu ^{^{\prime }},m,\kappa }
        [-\dot a_{\lambda ,\nu ^{^{\prime }}}^{*}a_{\lambda ,\nu }
        {\cal D}_{\mu,m}^{\lambda *}
        {\cal D}_{\mu ,\nu ^{^{\prime }}}^\lambda q_\kappa 
        (M_\kappa)_{\nu ,m}+ \dot a_{\lambda ,\nu }
        a_{\lambda ,\nu ^{^{\prime }}}^{*}
        {\cal D}_{\mu ,\nu }^{\lambda *}
        {\cal D}_{\mu ,m}^\lambda q_\kappa (M_\kappa)_{m,\nu }]\,. \nonumber
\end{eqnarray}

Using the unitarity condition for the ${\cal D}_{\mu ,\nu }^\lambda
(\omega _i)$ functions [$\sum_\mu {\cal D}_{\mu ,\nu }^{\lambda *}(\omega
_i){\cal D} _{\mu ,\nu ^{^{\prime }}}^\lambda (\omega _i)=\delta _{\nu
,\nu ^{^{\prime }}}$], the vibrational part of the kinetic energy
simplifies to

\begin{equation}
{\cal T}_{vibration}$=$\displaystyle{\frac B2}\sum_\nu \dot a_{\lambda ,\nu
}^{*}\dot a_{\lambda ,\nu } \,,
\end{equation}

\noindent
while the rotational part is

\begin{eqnarray}
{\cal T}_{rotation}&=&\displaystyle{\frac B2}\sum_{\kappa ,\kappa ^{^{\prime
}}}q_\kappa q_{\kappa ^{^{\prime }}}\sum_{\nu ,\nu ^{^{\prime }}}a_{\lambda
,\nu }a_{\lambda ,\nu ^{^{\prime }}}^{*}(M_\kappa M_{\kappa ^{^{\prime
}}})_{\nu ,\nu ^{^{\prime }}}  \nonumber \\
&=&\displaystyle{\frac 12}\sum_{\kappa ,\kappa ^{^{\prime }}}q_\kappa
q_{\kappa ^{^{\prime }}}{\cal J}_{\kappa \kappa ^{^{\prime }}} \,,
\end{eqnarray}

\noindent
where the inertia tensor is defined as

\begin{equation}
{\cal J}_{\kappa \kappa ^{^{\prime }}}=B\sum_{\nu ,\nu ^{^{\prime
}}}a_{\lambda ,\nu }a_{\lambda ,\nu ^{^{\prime }}}^{*}(M_\kappa M_{\kappa
^{^{\prime }}})_{\nu ,\nu ^{^{\prime }}}.
\end{equation}

\noindent
The coupling between the internal and rotational degrees of freedom
maintains a complicated structure, i.e.

\begin{eqnarray}
{\cal T}_{coupling} &=&i\displaystyle{\frac B2}\sum_{\nu ,\nu ^{^{\prime
}},\kappa }q_\kappa [\dot a_{\lambda ,\nu }a_{\lambda ,\nu ^{^{\prime
}}}^{*}-\dot a_{\lambda ,\nu ^{^{\prime }}}^{*}a_{\lambda ,\nu }](M_\kappa
)_{\nu ,\nu ^{^{\prime }}}  \nonumber \\
&=&i\sqrt{21}B\left\{ \vec{q}\left\{ a_3 \dot a_3 \right\} ^1\right\} _0^0\,,
\end{eqnarray}

\noindent
where $\left\{ {}\right\} _M^J$ means the angular momentum coupling to
$J,M$, and the last line corresponds to the $\lambda=3$ case.

% ....................................................................
\subsection{Definition of the intrinsic system}

We now proceed to the definition of the intrinsic system. We have
isolated three rotational variables (the Euler angles), but we also
have ($2\lambda +1)$ coordinates $a_{\lambda ,\nu }$. It is necessary
to reduce their number to ($2\lambda -2)$. This is customary achieved by
imposing the condition

\begin{equation}
{\cal J}_{\kappa \kappa ^{^{\prime }}}={\cal J}_\kappa \delta _{\kappa
\kappa ^{^{\prime }}}\,.
\end{equation}

\noindent
In fact, the requirement is that the off-diagonal elements of ${\cal J}
_{\kappa \kappa ^{^{\prime }}}$ vanish. These three conditions, which
usually are non-lineal, in principle reduce by three the number of
parameters. The inertia tensor ${\cal J}_{\kappa \kappa ^{^{\prime }}}$
can be expressed in terms of the coupling to angular momentum two and zero
of the collective variables, using the fact that

\begin{equation}
(M_\mu M_{\mu ^{^{\prime }}})_{\nu ,\nu ^{^{\prime }}}=\sum_J<1\mu 1\mu
^{^{\prime }}|JM>\left\{ MM\right\} _M^J\,.
\end{equation}

Due to symmetry reasons, the only values of $J$ that survive in the
summation related with the inertia tensor are two or zero. For $\lambda
=3$ the explicit expressions for the diagonal components of the inertia
tensor are the following: 

\begin{eqnarray}
{\cal J}_1&=&B\left[ 4\sqrt{7}\left\{ aa\right\} _0^0+\sqrt{21}\left\{
aa\right\} _0^2-3\sqrt{14}\Re\left\{ aa\right\} _2^2\right]  \nonumber \\
{\cal J}_2&=&B\left[ 4\sqrt{7}\left\{ aa\right\} _0^0+\sqrt{21}\left\{
aa\right\} _0^2+3\sqrt{14}\Re\left\{ aa\right\} _2^2\right]  \nonumber \\
{\cal J}_3&=&B\left[ 4\sqrt{7}\left\{ aa\right\} _0^0-2\sqrt{21}\left\{
aa\right\} _0^2\right]\,,
\end{eqnarray}

\noindent
while the vanishing of the off-diagonal components 
yields\footnote{
In the quadrupole case ${\cal J}_{12}$ is related to $\Im\left\{
aa\right\} _2^2$ while ${\cal J}_{13}$ and ${\cal J}_{23}$ are related to
$\left\{ aa\right\} _1^2$. The diagonal components of the inertia tensor
are related to $\left\{ aa\right\} _0^0$, $\left\{aa\right\} _0^2$ and
$\Re\left\{ aa\right\} _2^2$. In a similar way in the T = 1 pairing
collective description the off diagonal components of the inertia tensor
can be written as ${\cal J} _{12}=-B \Im \left\{ \Delta \Delta^*
\right\}^2_2$, ${\cal J}_{13}=-B \Re \left\{ \Delta \Delta^*
\right\}^2_1$, ${\cal J}_{23}=-B \Im \left\{ \Delta \Delta^*
\right\}^2_1$. The diagonal components are related to $\left\{ \Delta
\Delta^* \right\}^2_0$, $\Re \left\{ \Delta \Delta^* \right\}^2_2$ and $
\left\{ \Delta \Delta^* \right\}^0_0$}

\begin{eqnarray}
{\cal J}_{12}&=&3B\sqrt{14}\Im\left\{ aa\right\} _2^2=0  \nonumber \\
{\cal J}_{13}&=&3B\sqrt{14}\Re\left\{ aa\right\} _1^2=0  \nonumber \\
{\cal J}_{23}&=&3B\sqrt{14}\Im\left\{ aa\right\} _1^2=0.
\end{eqnarray}

It is convenient to define the auxiliary variables $u_m=\left\{ aa\right\}
_m^2.$\ The definition of the intrinsic system through the cancelation of
the off-diagonal components of the inertia tensor is therefore equivalent
to the conditions $u_1=0$, $u_2$ and $u_{-2}$ real. It must be noted that
similar conditions are held in the quadrupole case. In this case those
conditions are satisfied if $a_{2,1}=a_{2,-1}=0$ and a$ _{2,2}=a_{2,-2} $
and real. In the octupole case, one must satisfy three non-linear
equations, whose solutions are more complicated. 

The $a_{3\nu }$ (in what follows we will concentrate on the
$\lambda=3$ case, and therefore use $a_\nu $) must be
expressed in terms of four independent variables. Calling them $X,Y,Z$ and
$\gamma $ it is possible to use as a parametrization

\begin{eqnarray}
a_3&=&[\cos \gamma -\frac{\sqrt{3}}2\sin \gamma ]X+i[\cos \gamma +\frac{
\sqrt{3}}2\sin \gamma ]Y  \nonumber \\
a_2&=&\frac 1{\sqrt{2}}\sin \gamma Z \nonumber \\
a_1&=&\frac{\sqrt{5}}2\sin \gamma [X+iY] \nonumber \\
a_0&=&\sqrt{5}\cos \gamma Z 
\label{param}
\end{eqnarray}

\noindent
that automatically makes the off--diagonal components of the inertia
tensor equal to zero. 

To simplify the notation we define $x_\kappa =X,Y,Z$ and $\gamma _\kappa
=\gamma -\frac \pi 3\kappa $ for $\kappa =1,2,3.$
In terms of these variables, the $u_m$ different from zero can be written
as

\begin{eqnarray}
u_0&=&\eta \cos \gamma  \nonumber \\
u_2&=&\frac \eta {\sqrt{2}}\sin \gamma \,,
\end{eqnarray}

\noindent
where $\eta =5\sqrt{\frac 2{21}}\sum_\kappa x_\kappa ^2\cos \gamma
_\kappa$.

If we define

\begin{equation}
\rho ^2 \equiv \sum_\nu |a_\nu |^2=
\sum_\kappa x_\kappa ^2[1+4\cos ^2\gamma _\kappa ],
\end{equation}

\noindent
the diagonal components of the inertia tensor can be written as

\begin{equation}
{\cal J}_\kappa =B\left\{ 4\rho ^2-2\sqrt{21}\eta \cos \gamma _\kappa
\right\} .
\label{inertia}
\end{equation}

% ....................................................................
\subsection{Quantization of the Hamiltonian}

The next problem is that of quantizing the Hamiltonian. There is no
unique way to perform this quantification but, as in the quadrupole
case\cite{KB}, the Pauli prescription can be used. This recipe is designed
to give the right answer when the generalized variables are transformed to
Cartesian coordinates. Given a classical Hamiltonian written in terms of
variables $\alpha _\nu$ and their time derivatives, with a kinetic
energy given by

\begin{equation}
{\cal T=}\frac 12\sum_{mn}G_{mn}(\alpha )\dot \alpha _m\dot \alpha _n \,,
\end{equation}

\noindent
the Pauli prescription replaces the kinetic energy by the operator

\begin{equation}
{\cal T=-}\frac 12\sum_{mn}|G|^{-\frac 12}\frac \partial {\partial \alpha _m}
|G|^{\frac 12}G^{mn}\frac \partial {\partial \alpha _n} \,,
\end{equation}

\noindent
where $|G|$ is the determinant of the inertia matrix $G_{mn}$ and $G^{mn}$
is its inverse matrix. The volume element to be used is given by

\begin{equation}
\label{dvol}
d\tau =|G|^{\frac 12}\prod_nd\alpha _n \,.
\end{equation}

\noindent
In our case, the variables are $X,Y,Z,\gamma ,q_{1,}q_2,q_3$, and the
inertia matrix $G_{mn}$ is shown in Table \ref{table1}.

% -------------------------------------------------------------------
\section{Symmetry properties}

\subsection{General structure of the wave functions}

The total collective Hamiltonian will contain a potential energy depending
on the internal collective variables (and its time derivatives) in
addition to the kinetic energy terms already discussed. The eigenfunctions
of the total Hamiltonian can be labeled by the parity, the angular
momentum and its projection on the laboratory z-axis ($\tau ,I$ and
$I_z$). They can be conveniently expresses as a linear combination

\begin{equation}
\Psi _{I,I_z}^\tau (X,Y,Z,\gamma ,\omega _i)=\sum_Kg_{K,\xi }^\tau
(X,Y,Z,\gamma ){\cal D}_{I_z,K}^\lambda (\omega _i) \,,
\end{equation}

\noindent
where the quantum numbers $\xi$, related to the internal variables,
remain yet unspecified. This set of eigenfunctions constitute a complete
set of states and they will be orthogonal using $d \tau$ specified in
Eq.\ (\ref{dvol}) as volume element.

% \begin{equation}
% d\tau =|G|^{\frac 12}\prod_nd\alpha _n
% \end{equation}

% ....................................................................
\subsection{Invariants and the potential energy in Bohr's Hamiltonian}

As in the collective descriptions studied before\cite{BM2,KB,pair2}, there
are two types of symmetries that arise in the problem of collective
motion. The first type corresponds to the invariance under rotations. The
second type is related to the fact that even if the collective variables
in the Lab. system are uniquely determined, there are 24 different ways of
defining a right handed intrinsic system\cite{Bo52}. For simplicity, it
is commonly assumed that the potential energy does not depend on the
velocities, and that it should be an analytic function when 
written in terms of the $\alpha _\mu$. Naturally, it must also be a
scalar under rotations, and it is therefore important to search for simple
polynomials in terms of the $\alpha _\mu $ that are scalar under
rotations. It is well known that in the quadrupole case there are two of
the so called basic invariants\cite{KB} ($\beta ^2$ and $\beta ^3\cos
3\gamma $)  while in the T = 1 pairing case there are three basic scalars
under rotations [$\Delta ^2$ and $\Delta ^2e^{\pm 4i\phi }\cos
(2\Gamma )$], but only two if one requires invariance with regard to
rotations both in usual space and in gauge space [$\Delta ^2$ and
$\Delta ^4\cos ^2(2\Gamma )$]. 

In the octupole case it is possible to find in a simple way some of these
basic scalars with regard to rotations (we use the same notation as Kumar
and Baranger\cite{KB}). The first one is quite trivial (the upper index
denote the number of bosons needed to construct the scalar):

\begin{equation}
I^{(2)}=\rho ^2=\sum_\mu 
\alpha _\mu ^{*}\alpha _\mu =-\sqrt{7}\left\{ aa\right\} _0^0.
\end{equation}

\noindent
The second and third ones can be constructed in a similar way as in the
quadrupole case using $u_m$, the quadrupole variables related to the
inertia tensor, i.e. 

\begin{eqnarray}
I^{(4)}&=&\left\{ uu\right\} _0^0=\frac 1{\sqrt{5}}\eta ^2 \\
I^{(6)}&=&\left\{ \left\{ uu\right\} ^2u\right\} _0^0=-\sqrt{\frac 2{35}}
\eta ^3\cos 3\gamma \,.
\end{eqnarray}

To be sure that one has obtained all the basic scalars it is convenient to
count the states with angular momentum zero and check that they can be
constructed out of the elementary scalars already known. The number of
states having angular momentum zero for each number of bosons can be
constructed in a simple way using the m-scheme. Studying the states
obtained considering up to 40 bosons it is found that it is necessary to
introduce a basic scalar formed with ten bosons\footnote{We would like
thank to Prof. J. Blomqvist who called our attention to the existence of
this tenth order invariant.}. Its existence shows how much more
complicated is the octupole case as compared with the other collective
treatments already performed. 

% ......................................................................
\subsection{Wave function's symmetry properties }

In this subsection we will follow closely Ref.\cite{KB}. The wave function
$\Psi $ must be an analytical function of the coordinates $\alpha _\mu
$. The Lab. variables are unambiguously defined while, as Bohr already
noted, the intrinsic variables $a_m$ are not, as there are 24 ways of
defining a right handed intrinsic system starting from one Lab. system.

This can be seen more clearly if one defines three basic operators which
can be used to transform a given intrinsic system in an equivalent one:

\begin{itemize}
\item
${\cal R}_1:$ rotation through $\pi$ around the 1-axis of the
intrinsic system. 
\item
${\cal R}_2$ : rotation through $\frac \pi 2$ around the 3-axis
of the intrinsic system. 
\item
${\cal R}_3$ : cyclic permutation of the three intrinsic axes. 
\end{itemize}

Since ${\cal R}_1^2$ =${\cal R}_2^4={\cal R}_3^3=1$, the 24 possible
transformations between equivalent intrinsic systems are 
${\cal S} (s_1,s_2,s_3) = {\cal R}_1^{s_1} {\cal R}_2^{s_2}{\cal 
  R}_3^{s_3}$ with $0\leq s_1\leq 1$, $0\leq s_1\leq 3$, and $0\leq s_1\leq 2.$
There are then 24 ways of choosing the $a_m$ for a given $\alpha
_\mu$, but $\Psi $ must be the 
same for each of the choices, being invariant under the transformation
that changes an intrinsic system to an equivalent one, i.e. $\Psi $ must
be invariant under any of the transformations ${\cal S} (s_1,s_2,s_3)$.
All these transformations can be considered as acting on the coordinates $
(X,Y,Z,\gamma ,\theta ,\phi ,\psi )$. We have therefore that these
transformations applied to a point

\begin{itemize}
\item
${\cal R}_1 (X,Y,Z,\gamma ,\theta ,\phi ,\psi )=(X,Y,Z,\gamma
,\pi -\theta ,\phi +\pi ,-\varphi )$ 
\item
${\cal R}_2 (X,Y,Z,\gamma ,\theta ,\phi ,\psi )=( X,Y,Z,\gamma
,\theta ,\phi ,\varphi +\frac \pi 2)$ 
\item
${\cal R}_3 ( X, Y, Z, \gamma , \theta ,\phi ,\psi )=(
 X, Y, Z, \gamma$, three new Euler angles very complicated$)$ 
\end{itemize}

These very complicated new Euler angles related to ${\cal R}_3$ correspond
to the cyclic permutation of the axis but we will not need their explicit
form.

In order to obtain the corresponding transformation $R_k$ of wave
functions $\Psi $ we must apply the recipe for active transformations: 

\begin{center} 
new wave function at new point = old wave function at old point.
\end{center}

It is convenient to remember that the application of an active rotation
$R$ on a spherical tensor ${\cal T}_{lm}$ yields

\begin{equation}
{\cal [}R(\theta ,\phi ,\psi ){\cal T}]_{lm}=\sum_n{\cal D}_{m,n}^l(\theta
,\phi ,\psi ){\cal T}_{\ln }
\end{equation}

Writing the new wave function at the new point in terms of the old one at
the old point one obtains

\begin{equation}
R\,|I,M\rangle = \sum_K {\cal D}_{M,K}^I(R^{-1})\,|I,K\rangle = 
\sum_K[{\cal D}_{K,M}^I(\theta ,\phi ,\psi )]^{*} \, 
|I,K\rangle  \label{beta}
\end{equation}

\noindent
a similar relation holds between $a_m$ and $\alpha _\mu $ , i.e. $a_m$ $
=\sum_\mu [{\cal D}_{m,\mu }^3(\theta ,\phi ,\psi )]^{*}\alpha _\mu $. 

To obtain the effect of $R_i$ on the ${\cal D}_{K,M}^l$ matrices it is
convenient to relate $R_i$ to the operator $R_i^{\prime }$ that performs
the same transformation on the Lab. system, i.e. {\normalsize $R_i$ $
=RR_i^{\prime }R^{-1}$}, where $R$ corresponds to 
the rotation $(\theta ,\phi ,\psi)$. It is then possible to express

\begin{equation}
R_i{\cal D}_{M,N}^l(\theta ,\phi ,\psi )=RR_i^{\prime }R^{-1}{\cal D}
_{M,N}^l(\theta ,\phi ,\psi ) \,.
\end{equation}

\noindent
Applying three times Eq.\ (\ref{beta}) on the right hand side one gets

\begin{equation}
\sum_{KPQ}{\cal D}_{M,K}^l(R){\cal D}_{K,P}^l(R_i^{\prime -1}){\cal D}
_{P,Q}^l(R^{-1}){\cal D}_{Q,N}^l(\theta ,\phi ,\psi )=\sum_K{\cal D}
_{M,K}^l(\theta ,\phi ,\psi ){\cal D}_{K,N}^l(R_i^{\prime -1}) \,,
\end{equation}

\noindent
where in the last step the unitarity of the ${\cal D}_{K,N}^l$
matrices was used.

Taking into account that the wave functions can be written as

\begin{equation}
\Psi _{I,M}^\tau (X,Y,Z,\gamma ,\theta ,\phi ,\psi )=\sum_Kg_{I,K}^\tau
(X,Y,Z,\gamma ){\cal D}_{M,K}^\lambda (\theta ,\phi ,\psi ) \,,
\end{equation}

\noindent
and taking into account that the symmetry of the wave function with regard
to $ R_1,R_2$ and $R_3$ implies that

\begin{equation}
R_i \Psi _{I,M}^\tau (X,Y,Z,\gamma ,\theta ,\phi ,\psi)=\Psi _{I,M}^\tau
(X,Y,Z,\gamma ,\theta ,\phi ,\psi) \,,
\end{equation}

\noindent
and considering the effect of $R_i$ on the intrinsic and rotational
degrees of freedom 

\begin{eqnarray}
R_i{\cal D}_{M,K}^I(\theta ,\phi ,\psi )&=&{\cal D}_{M,K}^I(\theta ^{\prime
},\phi ^{\prime },\psi ^{\prime })=\sum_{K^{\prime }}{\cal D}_{M,K^{\prime
}}^I(\theta ,\phi ,\psi ){\cal D}_{K^{\prime },K}^I(R_i^{\prime -1}) \\
R_ig_{I,K}^\tau (X,Y,Z,\gamma )&=&g_{I,K}^\tau (X^{\prime },Y^{\prime
},Z^{\prime },\gamma ^{\prime }) \,,
\end{eqnarray}

\noindent
it is possible to obtain a condition on the part of the wave function that
depends on the internal variables:

\begin{equation}
g_{I,K^{\prime }}^\tau (X,Y,Z,\gamma )=\sum_Kg_{I,K}^\tau
(X^{\prime},Y^{\prime},Z^{\prime},\gamma ){\cal D}_{K^{\prime
},K}^I(R_i^{\prime -1}).  \label{di} 
\end{equation}

In order to obtain the effect of the symmetry operation on the internal
variables, we must first evaluate ${\cal D} _{M,K}^I(R_i)$ explicitly.
A lengthy yet straightforward calculation yields

\begin{eqnarray}
{\cal D}_{M,K}^I(R_1)&=&(-1)^I\delta _{M,-K} \\
{\cal D}_{M,K}^I(R_2)&=&(-i)^K\delta _{M,K} \\
{\cal D}_{M,K}^I(R_3)&=&{\cal D}_{M,K}^I(\frac \pi 2,0,\frac \pi 2) \,.
\end{eqnarray}

\noindent
This last matrix is shown in Table \ref{table2}.

From the transformation properties of the coefficients $a_m$, one can
deduce the transformation properties for $X,Y,Z$ and $\gamma$. In Table
\ref{table3} we show all the information related to the transformation
properties of the variables, as well as the characteristics of the matrices
related with ${\cal D}_{K,K^{\prime }}^I(R_i^{-1}).$

With regard to the spatial inversion ${\cal P}$, which is not a symmetry
property due to ambiguities in the definition of the intrinsic axis, but
must be a symmetry of the intrinsic wave function, one obtains ${\cal P}
:(X,Y,Z,\gamma )\rightarrow (-X,-Y,-Z,\gamma )$ and 
${\cal P}_{K,K^{\prime }}^I=(-1)^I\delta _{K,K^{\prime }}$ 
that replaces the ${\cal D} _{K,K^{\prime }}^I.$ in Eq.(\ref{di}). 
The different transformations will then yield some symmetry properties of
the wave function: 

\begin{eqnarray}
R_1:g_{I,K}^\tau (X,Y,Z,\gamma )&=&(-1)^Ig_{I,-K}^\tau
(X,-Y,-Z,\gamma ) \\
R_2:g_{I,K}^\tau (X,Y,Z,\gamma )&=& (-i)^{K} g_{I,K}^\tau
(-Y,X,Z,-\gamma ) \\
R_3:g_{I,K}^\tau (X,Y,Z,\gamma )&=&\sum_{K^{\prime
}}g_{I,K^{\prime }}^\tau (X,Y,Z,\gamma +\frac{2\pi }3){\cal D}_{K,K^{\prime
}}^I(\frac \pi 2,\frac \pi 2,\pi ).
\end{eqnarray}

There are some combinations of these operators that yield more useful
results, such as

\begin{eqnarray}
R_2^2:g_{I,K}^\tau (X,Y,Z,\gamma )&=&(-1)^Kg_{I,K}^\tau (-X,-Y,Z,\gamma ) \\
R_2^2R_1:g_{I,K^{\prime }}^\tau (X,Y,Z,\gamma )&=&(-1)^{I+K}g_{I,-K}^\tau
(-X,Y,-Z,\gamma ) \,,
\end{eqnarray}

\noindent
If the wave function is also an eigenfunction of the parity operator with
eigenvalue $\Pi$, i.e. 

\begin{equation}
{\cal P}\Psi _{I,M}^\tau (X,Y,Z,\gamma ,\theta ,\phi ,\psi )=\Pi \Psi
_{I,M}^\tau (X,Y,Z,\gamma ,\theta ,\phi ,\psi) \,,
\end{equation}

\noindent
we obtain that 

\begin{equation}
{\cal P}:g_{I,K}^\tau (X,Y,Z,\gamma )=\Pi (-1)^Ig_{I,K}^\tau
(-X,-Y,-Z,\gamma ) \,.
\end{equation}

% -----------------------------------------------------------------------
\section{The rigid rotor }

In this section we study the axially symmetric rotor. Axial symmetry
implies that at least two of the diagonal components of the inertia tensor
must be equal. From Eq. (\ref{inertia}) it follows that for $\gamma=0$ or
$\gamma=\frac \pi 3$ this condition is satisfied. It is also possible to
make the three components of the inertia tensor equal by making all the
three $x_k$ equal (in this particular case $\eta=0$). For $\gamma=0$
one has ${\cal J}_1={\cal J}_2$ while for $\gamma=\frac \pi 3$ one has
${\cal J}_1={\cal J}_3$. In both cases the remaining component does not
vanish. We will study in some detail the $\gamma=0$ case because it
illustrates some of the complications and features related to the octupole
degrees of freedom.  

We will assume in this section that the $\gamma$
degree of freedom is frozen (in particular $\gamma$ is equal to zero). It
follows also from the relation between the collective variables in the
intrinsic system ($a_\mu$) and the intrinsic variables ($X,Y,Z$ and
$\gamma$) given by Eqs. (\ref{param}) that $\gamma=0$ implies that the
only deformations allowed should have $K=0$ or $K=\pm 3,\;\pm 6,...$ 
It must be noted
that when the collective description of the T = 1 pairing excitations was
studied in Refs. \onlinecite{pair2,pair3}, it was found that for studying the
rigid rotor it was necessary to retain the gauge angle degree of freedom
as the rotational Hamiltonian was coupled to it. In this section, even if
the rotational Hamiltonian is coupled to the $\gamma$ degree of freedom we
will assume that it is possible to freeze it and we will therefore
disregard it completely. In this case the two basic quantities
$\rho^2$ and $\beta$ can be written as: 
 
\begin{eqnarray}
\rho^2&=&2(X^2+Y^2)+5Z^2, \\
\beta&=&\frac{10}{\sqrt{21}}(-\frac1 2(X^2+Y^2)+Z^2) \,,
\end{eqnarray}

\noindent
and the inertia tensor has a rather simple expression in terms of the
intrinsic variables:

\begin{eqnarray}
{\cal J}_1 &=&{\cal J}_2=3(X^2+Y^2)+30Z^2 \nonumber \\
{\cal J}_3&=&18(X^2+Y^2) \,.
\end{eqnarray}

It is convenient to reparametrize the problem in order to introduce $\rho$
as one of the dynamical variables and use ``spherical coordinates'' by
defining two angle variables $\mu$ and $\zeta$

\begin{eqnarray}
X&=&\rho \sin\mu \cos\zeta /\sqrt2,  \nonumber \\
Y&=&\rho \sin\mu \sin\zeta /\sqrt2,  \nonumber \\
Z&=&\rho \cos\mu /\sqrt5.
\end{eqnarray}

\noindent
In terms of these variables $\beta =\frac{10}{\sqrt{21}}\rho ^2[-\frac 14
\sin^2\mu +\frac 15\cos^2\mu ]$. Now the $G_{mn}$ inertia matrix will be a
six by six matrix as shown in Table \ref{table4}. 

We will assume the adiabatic hypothesis for $\rho $ and $\mu$. But it must
be noted that, as $\zeta $ is a cyclic coordinate, we cannot use for it the
adiabatic hypothesis and we must, therefore, consider the coupling of
$\dot\zeta$ with $q_1,q_2$ and $q_3$ in an exact way. For this purpose
we use the conjugate momentum method, i.e., we introduce

\begin{eqnarray}
Q_1&=&\frac{\partial {\cal T}}{\partial q_1} =
  6\rho^2[1-\frac{3}{4} \sin^2\mu]q_1   \\
Q_2&=&\frac{\partial {\cal T}}{\partial q_2} =
  6\rho^2[1-\frac{3}{4} \sin^2\mu]q_2   \\
Q_3&=&\frac{\partial {\cal T}}{\partial q_3} = 
  9\rho^2\sin^2\mu [q_3-\frac{1}{3} \dot\zeta]=K  
\label{K3} \\
P_\zeta&=&\frac{\partial {\cal T}}{\partial \dot\zeta} = 
  -3\rho^2 \sin^2\mu[q_3-\frac{1}{3} \dot\zeta] = 
  -i\frac{\partial}{\partial\zeta} = m_\zeta \,.
\label{mk}
\end{eqnarray}

\noindent
It must be noted that $Q_i$ satisfy the commutation relations given by
Eqs.(\ref{conmu}). Besides, the relation between Eq.(\ref{K3}) and Eq.
(\ref{mk}) provides the auxiliary condition: 

\begin{equation}
K=-3m_\zeta=0,\pm3,\pm6,...
\end{equation}

The kinetic energy can be written in terms of these new variables
$Q_{\kappa}$ as

\begin{equation}
{\cal T}_{rot}=\sum_{\kappa} {\frac{1}{2{\cal J}_{\kappa}}}
Q^2_{\kappa} \,,
\end{equation}

\noindent
where the components of the moment of inertia are given by

\begin{eqnarray}
{\cal J}_1 &=&{\cal J}_2=\frac 3 4 \rho^2 [1+3\cos^2\mu] \nonumber\\
{\cal J}_3&=&\frac{9}{2}\rho^2 \sin^2\mu \,.
\end{eqnarray}

The total rotational energy can be written as

\begin{equation}
{\cal T}_{rot}=\frac {I(I+1)} {2{\cal J}_1} + 
\frac {K^2 ({\cal J}_1-{\cal J}_3)} {2 {\cal J}_1 {\cal J}_3}=
\displaystyle{\frac{[I(I+1)+\frac 2 3 K^2 \frac{[1-
\frac 9 4 \sin^2 \mu]}{\sin^2\mu}]}{2{\cal J}_1}}  \,.
\label{ener}
\end{equation}

\begin{figure}
\begin{center}
\leavevmode
\epsfbox{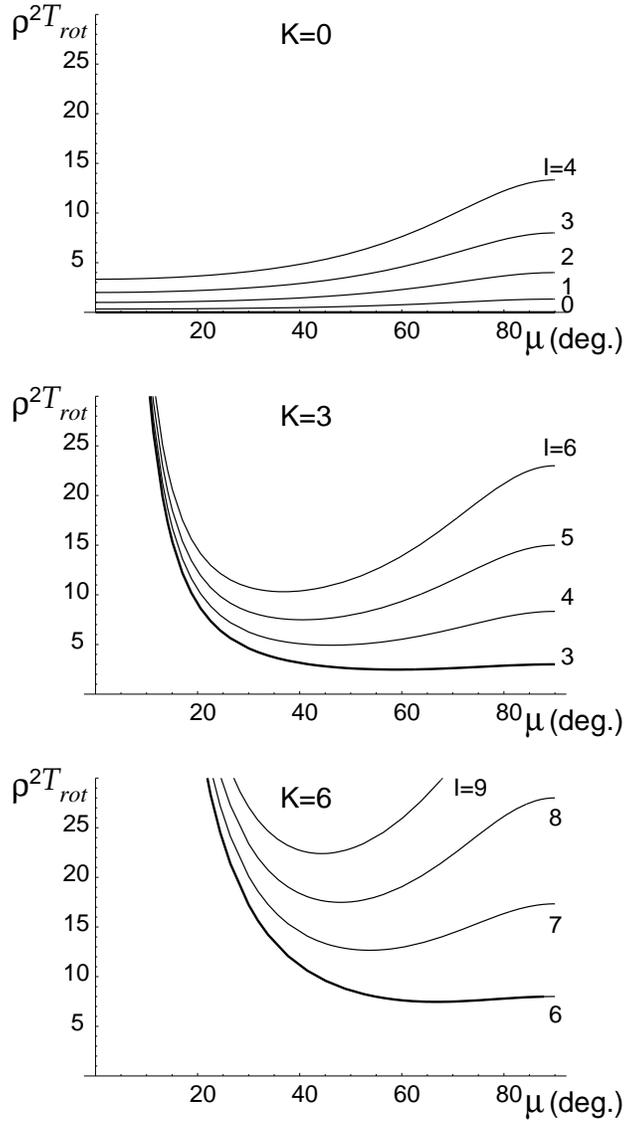}
\end{center}
\caption{ \label{fig:spect}
Spectrum of a symmetric ($\gamma=0$) rigid rotor for $K=0,3,6$ as a
function of the deformation parameter $\mu$.
}
\end{figure}

\noindent
The structure of the full ``rotational band'' is now more complicated than
for the quadrupole case. The part of the band that has $K=0$ has a
structure similar to that of the usual rigid quadrupole band, 
but the part of the
band related to the states with $K \ne 0$ has energies determined also by
$\rho$ and $\mu$ that are related with the energy of the first member of
the ground state rotational band. 
Note that $\mu$ represents deviations
from perfectly axisymmetric deformations, i.e. for $\mu=0$ the
deformations have ${\cal J}_3=0$ and therefore $K\equiv 0$, only for
$\mu \neq 0$ can nonzero $K$ be found. 
For $\mu=0$ the rotational band has the
usual structure (all states have $K=0$, as states with $K \ne 0$
have an infinite energy). When the states with angular momentum equal or
larger than $3$ are considered it is found that there are more than one
state belonging to the ground state band with this angular momentum: one
with $K=0$ and parity $(-1)^I$, and two with $K=3$ and energies given by Eq.
(\ref{ener}). Each time that $I=|K|= 3n$ (with $n$ integer) two new states
appear (i.e. we will have five states with $8 \ge I \ge 6$, seven states
with $11 \ge I \ge 9$, etc.).  
Figure \ref{fig:spect} shows the lowest energy states for $K=0,3,6$ as
a function of $\mu$.

% ----------------------------------------------------------------------
\section{Summary}

In the present paper we have obtained a parametrization of the octupole
collective variables that guarantees that the inertia tensor is diagonal. 
The relation between the components of the octupole intrinsic variables
and those defined parameters is given by the equations

$a_3=[\cos \gamma -\frac{\sqrt{3}}2\sin \gamma ]X+i[\cos \gamma
+\frac{\sqrt{ 3}}2\sin \gamma ]Y$

$a_2=\frac 1{\sqrt{2}}\sin \gamma Z$

$a_1=\frac{\sqrt{5}}2\sin \gamma [X+iY]$

$a_0=\sqrt{5}\cos \gamma Z$

In addition, we have identified the way in which the intrinsic
variables ($x_k$ and $\gamma$) transform with regard to the symmetry
operations $R_i$. These can be summarized as follows:  

a) The variables $x_k$ transform in a similar way as the coordinates of a
vector related to the intrinsic axis,

b) $\gamma $ transforms in the same way as in the quadrupole case. This is
not at all strange due to the particular relation of $\gamma $ to the
inertia tensor. 

\noindent
The first condition means that the parameters $X,Y$ and $Z$ can be
considered as positive, as all the other signs can be obtained by
relabeling the axis. The second statement, as in the quadrupole case, tell
us that all possible shapes are contained in the range $0 \le \gamma \le
\frac \pi 3$.

A general octupole field can be written in an alternative way\cite{ha} as
$V_3=\sum_ma_mY_{3m}^{*}=\epsilon _0A_2+\sum_{i=1}^3\epsilon _1(i)F_1(i)+$
$\sum_{i=1}^3\epsilon _2(i)F_2(i)$, where $A_1,F_1(k)$ and $F_2(k)$ are
related to the irreducible representations of the octahedron group. In
Ref.\cite{ha} it was proposed that an appropriate parametrization can be
obtained imposing the cancelation of $\epsilon _2(i)\forall i.$ Using the
parametrization that we are proposing, these parameters have now a rather
simple structure:

$\epsilon _0=0$

$\epsilon _1(1)=-\sin \left( \gamma \right) Z $

$\epsilon _1(2)=-\sin \left( \gamma -\frac{2\pi }3\right) X $

$\epsilon _1(3)=\sin \left( \gamma -\frac{4\pi }3\right) Y $

$\epsilon _2(1)=\sqrt{5}\cos \left( \gamma \right) Z $

$\epsilon _2(2)=\sqrt{5} \cos \left( \gamma -\frac{2\pi }3\right) X $

$\epsilon _2(2)=- \sqrt{5}\cos \left( \gamma \frac{4\pi }3\right) Y \,.$

It must be noted that, for the octupole degrees of freedom, the
existence of axial symmetry does not necessarily imply that the
rotational band must have K = 0. The study of the rigid rotor has
illustrated the richness of the octupole rotational bands. 

Last but not least, it is worthwhile noticing that if one takes into
account simultaneously the quadrupole and octupole degrees of freedom, and
for each one of them the condition that the inertia tensor has to be
diagonal is imposed, the intrinsic system for both degrees of freedom will
not be the same: each degree of freedom will have its own "intrinsic
system". It must be remembered that the Euler angles associated to each
intrinsic system are just a way of labeling three of the five (in the
quadrupole case) or seven (in the octupole one) collective dynamical
variables. 

Discussions with R. P. J. Perazzo and J. Fernandez Niello are gratefully
acknowledged. This work has been supported in part by the Carrera del
Investigador Cient\'\i fico y T\'ecnico, by PID $N^o4547/96$ of the
CONICET, Argentina, PMT-PICT1855 of ANPCYT and Grant Ex-055 from the
University of Buenos Aires. CW was supported in part by the 
University of Buenos Aires.

\vfill
\eject

\begin{table}
\caption{The inertia matrix $G_{m,n}$}
\begin{center}
\leavevmode
\epsfbox{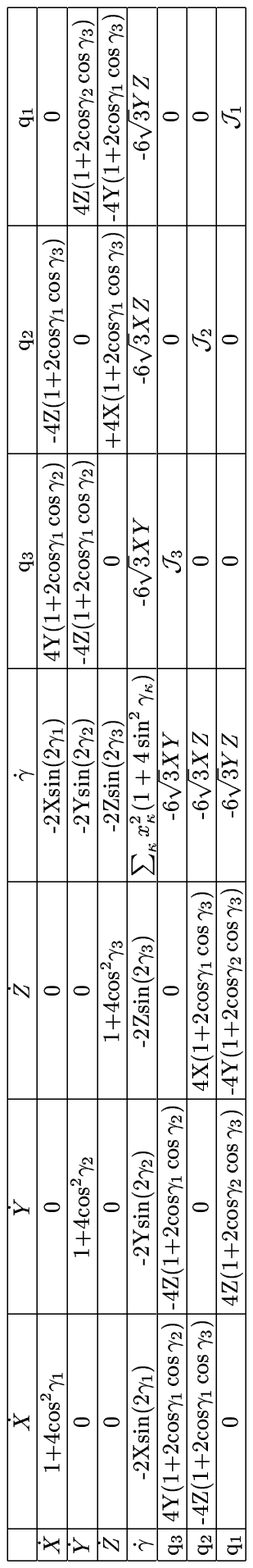}
\end{center}
\label{table1}
\end{table}

\begin{table}
\caption{Matrix elements of ${\cal D}^I_{M,K} (\frac \pi 2,0,\frac \pi 2)$}
\begin{tabular}{|c|c|c|c|c|c|c|c|}
$M\backslash K$ & $3$ & $2$ & $1$ & $0$ & $-1$ & $-2$ & $-3$ \\ 
\tableline
$3$ & $-\frac i8$ & $-i\frac{\sqrt{6}}8$ & $-i\frac{\sqrt{15}}8$ & $-i\frac{
\sqrt{20}}8$ & $-i\frac{\sqrt{15}}8$ & $-i\frac{\sqrt{6}}8$ & $-\frac i8$ \\ \hline
$2$ & $\frac{\sqrt{6}}8$ & $\frac 48$ & $\frac{\sqrt{10}}8$ & $0$ & $-\frac{
\sqrt{10}}8$ & $-\frac 48$ & $-\frac{\sqrt{6}}8$ \\ \hline
$1$ & $i\frac{\sqrt{15}}8$ & $i\frac{\sqrt{10}}8$ & $-\frac i8$ & $-i\frac{
\sqrt{12}}8$ & $-\frac i8$ & $i\frac{\sqrt{10}}8$ & $i\frac{\sqrt{15}}8$ \\ \hline
$0$ & $-\frac{\sqrt{20}}8$ & $0$ & $\frac{\sqrt{12}}8$ & $0$ & $-\frac{\sqrt{
12}}8$ & $0$ & $\frac{\sqrt{20}}8$ \\ \hline
$-1$ & -$i\frac{\sqrt{15}}8$ & $i\frac{\sqrt{10}}8$ & $\frac i8$ & $-i\frac{
\sqrt{12}}8$ & $\frac i8$ & $i\frac{\sqrt{10}}8$ & -$i\frac{\sqrt{15}}8$ \\ \hline
$-2$ & $\frac{\sqrt{6}}8$ & $-\frac 48$ & $\frac{\sqrt{10}}8$ & $0$ & $-
\frac{\sqrt{10}}8$ & $\frac 48$ & $-\frac{\sqrt{6}}8$ \\ \hline
$-3$ & $\frac i8$ & $-i\frac{\sqrt{6}}8$ & $i\frac{\sqrt{15}}8$ & $-i\frac{
\sqrt{20}}8$ & $i\frac{\sqrt{15}}8$ & $-i\frac{\sqrt{6}}8$ & $\frac i8$
\end{tabular}
\label{table2}
\end{table}

\begin{table}
\caption{Action of the different transformation operators on the 
corresponding intrinsic variables. Also shown are the explicit 
representation of  ${\cal D}^I_{K,K'} (R^{-1}_i)$}
\begin{tabular}{|c|c|c|c|}
$Variable\backslash Transformation$ &$ R_1$ & $R_2$ & $R_3$ \\ \tableline
X' & X & -Y & Z \\ \hline
Y' & -Y & X & X \\ \hline
Z' & -Z & Z & Y \\ \hline
$\gamma $ & $\gamma $ & -$\gamma $ & $\gamma +\frac{2\pi }3$ \\ \hline
$\theta $ & $\pi -\theta $ & $\theta $ & Complicated \\ \hline
$\phi $ & $\pi +\phi $ & $\phi $ & Complicated \\ \hline
$\psi $ & $-\psi $ & $\psi +\frac \pi 2$ & Complicated \\ \hline
${\cal D}_{K,K^{\prime }}^I(R_i^{-1}).$ & $(-1)^I\delta _{K,-K^{\prime }}$ & 
$(-i)^K\delta _{K,K^{\prime }}$ & ${\cal D}_{K,K^{\prime }}^I(\frac \pi 2,
\frac \pi 2,\pi ).$
\end{tabular}
\label{table3}
\end{table}

\begin{table}
\caption{Inertia matrix $G_{mn}$ for the $\gamma=0$ case}
\begin{tabular}{|c|c|c|c|c|c|c|}
& $\dot\rho$ & $\dot\mu $& $\dot\zeta $&$ q_1$ & $q_2$ & $q_3$ \\
\tableline 
$\dot\rho$ & 1 & 0 & 0 & 0 & 0 & 0 \\ \hline
$\dot\mu$ & 0 & $\rho ^2 $& 0 & 0 & 0 & 0 \\ \hline 
$\dot\zeta$ & 0 & 0 &$ \rho ^2\sin ^2\mu $& $-3\rho ^2\sin ^2\mu $& 0
& 0 \\ \hline 
$q_1$ & 0 & 0 & $-3\rho ^2\sin ^2\mu$ &$ 9\rho ^2\sin ^2\mu$ & 0 & 0
\\ \hline
$q_2$ & 0 & 0 & 0 & 0 & $6\rho ^2[1-\frac 34\sin ^2\mu ]$ & 0 \\ \hline
$q_3$ & 0 & 0 & 0 & 0 & 0 & $6\rho ^2[1-\frac 34\sin ^2\mu ]$
\end{tabular}
\label{table4}
\end{table}

\vfill
\eject

%==========================================================================

%=====================================================================
%=====================================================================
\end{document}